\documentclass[useAMS,usenatbib]{mn2e}
\usepackage{amsmath}
\usepackage{float}
\usepackage{graphicx}
\usepackage{txfonts}
\usepackage{indentfirst}
\usepackage{color}
\usepackage{ulem}
\usepackage{hyperref}
\newcommand{\eq}[1]{\begin{equation}  #1 \end{equation}}
\newcommand{\eqs}[1]{\begin{equation} \begin{split} #1 \end{split} \end{equation}}

\newcommand{\br}[1]{\left( #1 \right)}

\newcommand{\bb}[1]{\left[ #1 \right]}

\newcommand{\dd}{{\rm d}}

\newcommand{\msun}{\mbox{$\,{\rm M}_{\sun}$}}

\def\apj{ApJ}

\def\aap{A\&A}

\def\mnras{MNRAS}

\def\R2500c{\ensuremath{r_{\rm{2500c}}}}
\def\R500c{\ensuremath{r_{\rm{500c}}}}
\def\R500m{\ensuremath{r_{\rm{500m}}}}
\def\m500c{\ensuremath{M_{\rm{500c}}}}
\def\m500m{\ensuremath{M_{\rm{500m}}}}

\def\MHSE{\ensuremath{M_{\rm HSE}}}
\def\Mcorr{\ensuremath{M_{\rm corr}}}
\def\Mtrue{\ensuremath{M_{\rm true}}}

\title[Removing the HSE mass bias]{Analytical model
for non-thermal pressure in galaxy clusters - III. Removing
the hydrostatic mass bias}

\author[Xun Shi, Eiichiro Komatsu,
Daisuke Nagai and Erwin T. Lau]{Xun Shi$^{1}$\thanks{E-mail:
xun@mpa-garching.mpg.de}, Eiichiro Komatsu$^{1,2}$,
Daisuke Nagai$^{3,1}$, Erwin T. Lau$^{3}$ \\
$^{1}$Max-Planck-Institut f\"ur Astrophysik,
Karl-Schwarzschild-Stra{\ss}e 1, D-85740 Garching bei M\"unchen, Germany\\
$^{2}$Kavli Institute for the Physics and
Mathematics of the Universe (Kavli IPMU, WPI), Todai Institutes for Advanced Study, the
University of Tokyo,\\ Kashiwa 277-8583, Japan\\
$^{3}$Department of Physics, Yale University, New Haven, CT 06520, U.S.A.
}

\begin{document}


\pagerange{\pageref{firstpage}--\pageref{lastpage}} 

\maketitle

\label{firstpage}

\begin{abstract}
 Non-thermal pressure in galaxy clusters leads to underestimation of the
 mass of galaxy clusters based on hydrostatic equilibrium with thermal
 gas pressure. This occurs even for dynamically relaxed clusters that are
 used for calibrating the mass-observable scaling relations.
 We show that the analytical model for non-thermal pressure developed in \citet{shi14} can correct for this
 so-called `hydrostatic mass bias', if most of the non-thermal
 pressure comes from bulk and turbulent motions of gas in
 the intracluster medium.
 Our correction works for the sample average irrespective of the
 mass estimation method, or the dynamical state of the clusters. This makes it
possible to correct for the bias in the hydrostatic mass estimates from
 X-ray surface brightness and the Sunyaev-Zel'dovich
 observations that will be available for clusters in a wide range of
 redshifts and dynamical states.
\end{abstract}

\begin{keywords}
galaxies: clusters: general -- galaxies: clusters: intracluster medium --
cosmology: observations -- methods: analytical -- methods: numerical
\end{keywords}

\section[]{Introduction}

Galaxy clusters promise great statistical power as a low redshift
cosmological probe, if their masses are estimated accurately
\citep{allen/evrard/mantz:2011}. A bias in the mass estimation
poses the most serious challenge to the success of cluster cosmology.
For example, the recently reported tension in cosmological inferences
from the cosmic microwave background (CMB) and galaxy clusters measured
by the Planck satellite, the so-called `Planck CMB-Cluster tension'
\citep{planck13a,planck15XXIV}, may point to new physics, but it may
also reflect the urgent need for better understanding and correction for
systematic biases \citep[see e.g.][]{vdl14,lei14,donahue14,israel15}.

The most prominent astrophysical bias has to do with the assumption
of hydrostatic equilibrium (HSE) of the intracluster medium (ICM) inside the
gravitational potential. HSE no longer holds in the presence of non-thermal
sources of pressure.
Hydrodynamical simulations show significant pressure associated with
bulk and turbulent motions of the gas, which leads to a
`HSE mass bias' of order ten percent or more in the mass estimates of individual
clusters
\citep[e.g.,][]{kay04,rasia06,rasia12,nagai07b,piff08,lau09,men10,nelson12,nelson14b}.

As the amplitude of non-thermal pressure can hardly be measured
directly by observations, one has to resort to theoretical
estimations to correct for this bias. To this end,
we have developed a physically motivated analytical model for
non-thermal pressure in the ICM \citep[][SK14 hereafter]{shi14}.
Given the accretion history of the
cluster, the model predicts the amplitude of non-thermal pressure
and its radial, mass, and redshift dependencies. The model successfully
reproduces the average non-thermal pressure profiles seen in
cosmological hydrodynamical simulations \citep{bat12}, the average
thermal pressure profile measured by Planck \citep{arnaud10}, as well as
non-thermal pressure profiles of {\it individual} clusters in
hydrodynamical simulations \citep[][SKNN15 hereafter]{shi15}.

In this paper, we study the ability of the SK14 model in correcting the HSE
mass estimation by testing it on simulated clusters in a set of high-resolution
cosmological hydrodynamical simulations (\citealt{nelson14b}), with special
attention paid to the dynamical state of galaxy clusters.

The dynamical state of a cluster influences its mass estimation in
two ways. First, the more dynamically relaxed a cluster is, the smaller its
non-thermal pressure (e.g. SK14; \citealt{nelson14}), and thus the smaller its
HSE mass bias. Second, the cluster profiles are smoother when a cluster is
dynamically more relaxed, which gives a better precision in the mass estimation.

In the early days, HSE mass estimates were applied to clusters irrespective
of their dynamical state \citep[e.g.][]{reiprich02}. These studies use X-ray
surface brightness data and make the assumption that the ICM is isothermal.
After the recognition that gas temperature decreases systematically with
increasing radius outside the cluster core region, precise HSE mass estimates
require also spatially-resolved, high signal-to-noise gas temperature
measurements, which are harder to obtain especially out to high
redshifts.

The current cluster mass estimates for a large sample of clusters
are mostly performed statistically (rather than
individually) using scaling relations between the mass
and spatially-averaged observables, such as the mean X-ray temperature,
luminosity, or their combination, and the Sunyaev-Zel'dovich (SZ)
effect \citep{sunyaev70,sunyaev72} averaged over certain cluster
radii (e.g. \citealt{vik09a,and11,bender14,planck14XX}, see also
\citealt{gio13} for a review). In this context, HSE mass estimates are
usually performed only on a small sample of dynamically relaxed clusters
where the bias from non-thermal pressure is minimized, and are used for
observationally calibrating the scaling relations with the hope that the adopted scaling relation is
robust against varying dynamical states (which holds true for some
scaling relations, see \citealt{krav06}).

However, even for relaxed clusters, non-thermal
pressure is expected to be non-negligible because non-thermal motions are
continuously sourced by the mass accretion and are never fully thermalized,
especially at the cluster outskirts (SK14; \citealt{nelson14}). Therefore, it is important to study the
correction of HSE mass bias induced by non-thermal
pressure for dynamically relaxed clusters. This is one of the primary goals of
this paper.

In recent years, the advances in SZ observations make it possible to
estimate HSE masses using X-ray surface brightness combined with the gas pressure profiles
probed by the SZ effect, without the expensive X-ray temperature
measurements.
This will make individual cluster mass estimates for a large number of
clusters possible out to high redshifts, and this time with high precision
to large radii thanks to the relative insensitivity of the SZ effect to gas
density.
In light of these observational advances, another goal of this paper is to
study the possibility of correcting HSE mass bias for a population of
clusters in a wide range of redshifts and dynamical states.

The rest of the paper is organized as follows: we introduce our method of
correcting non-thermal pressure in cluster mass estimation in
Sect.\;\ref{sec:mass_method}, where we discover the necessity of regulating the
input cluster profiles by fitting or smoothing. Then we present four
methods for regularizing the profiles in Sect.\;\ref{sec:smoothing} and
\ref{sec:HSEmethods} and apply them to our simulated clusters. After introducing
the way we select subsample of clusters based on their dynamical state in
Sect.\;\ref{sec:dynstate}, we present our results in Sect.\;\ref{sec:results}
and conclude in Sect.\;\ref{sec:conclusion}. In the two appendices we study the
effect of velocity anisotropy (Appendix\;\ref{sec:anisotropy}) and how an error on the pressure profile
propagates to an error on the mass (Appendix\;\ref{app:massbias}).

\section[]{HSE Mass estimate and its correction}
\label{sec:massbias}

\subsection{Method}
\label{sec:mass_method}

The assumption of HSE relates the thermal pressure
$P_{\rm th}$ that is observable from both X-ray and SZ observations to
the hydrostatic mass $M_{\rm HSE}$ as
\eq{
\label{eq:MHSE}
M_{\rm HSE}(<r) \equiv  - \frac{r^2}{G \rho_{\rm gas}(r)} \frac{\partial
  P_{\rm th}}{\partial r} \,,
}
where $\rho_{\rm gas}$ is the gas density.
Since the total pressure in the ICM, $P_{\rm tot} \equiv P_{\rm th} +
P_{\rm nth}$ where $P_{\rm nth}$ is non-thermal pressure support, is
greater than $P_{\rm th}$ by definition, $M_{\rm HSE}$ is an
underestimate of the true mass. We thus define the `corrected mass'
$M_{\rm corr}$ as\footnote{Unlike for $P_{\rm th}$, random motions producing
$P_{\rm nth}$ need not be isotropic. While we have implicitly assumed
the random motions to be isotropic here, we investigate the consequence
of velocity anisotropy in Appendix\;\ref{sec:anisotropy}.}
\eq{
\label{eq:Mcorr}
M_{\rm corr}(<r) \equiv  - \frac{r^2}{ G \rho_{\rm gas}(r)}
\frac{\partial P_{\rm tot}}{\partial r} \,.
}
This $M_{\rm corr}$ is expected to be close to an
unbiased estimate of the total mass when the non-thermal
pressure support is the dominant source of the HSE mass bias, as is the case
found in hydrodynamical simulations at least for regions within $r_{500}$
\citep[e.g.][]{lau13,nelson14}.

Cluster masses are usually defined with respect to a radius $r_{\Delta}$, within
which the mean density equals e.g. $\Delta = 500$ times the critical
density of the Universe $\rho_{\rm crit}$,
\eq{
\label{eq:mdelta}
M_{\Delta} \equiv \frac{4\pi r_{\Delta}^3}{3}  \rho_{\rm crit} \Delta  \,.
}

Solving for $M_{\Delta}$ using equation (\ref{eq:MHSE}) or
(\ref{eq:Mcorr}) is equivalent to finding the
corresponding radius $r_{\Delta}$ from a spherically-averaged `overdensity' profile defined by
\eq{
\label{eq:deltaprof}
\mathbf{\Delta}(r) \equiv \frac{1}{\rho_{\rm crit}} \frac{3 G M(<r)}{4\pi r^3} =
- \frac{3}{4 \pi \rho_{\rm crit}} \frac{1}{r \rho_{\rm gas}(r)} \frac{\partial
P}{\partial r}\,, } with the implicit relation $\mathbf{\Delta}(r_{\Delta}) =
\Delta$. Here, $P$ is either $P_{\rm th}$ or $P_{\rm tot}$.
An error on the pressure gradient due to neglecting $P_{\rm nth}$ will
lead to an error on $M_{\Delta}$ through an incorrect estimation of
$r_{\Delta}$.

\begin{figure}
\centering
\includegraphics[width=0.48\textwidth]{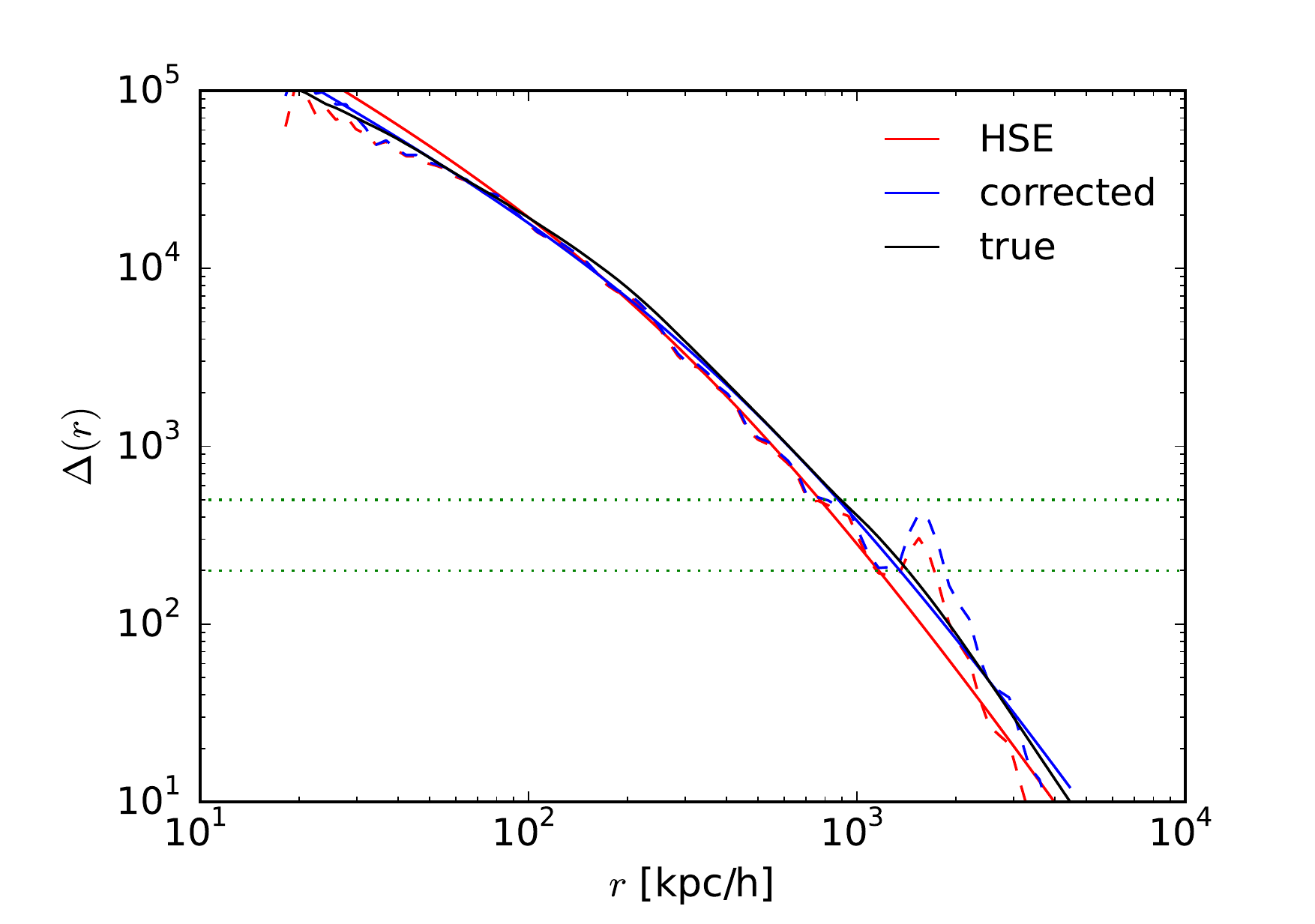}
\caption{Spherically-averaged overdensity profiles (equation
 \ref{eq:deltaprof}) with respect to the critical density,
 $\mathbf{\Delta}(r)$, of a simulated cluster. The horizontal dotted
 lines show $\Delta=500$ and 200. The true overdensity profile (black solid
 line) is calculated from the mass profile in the simulation, whereas the
 estimated profiles are calculated from
 spherically-averaged gas density and pressure profiles in the simulation using
 equations~(\ref{eq:MHSE}) and (\ref{eq:Mcorr}) (red and blue lines,
 respectively).
 Structures and noises in the gas density and pressure profiles are greatly
 amplified by the derivative and division procedures required in
 computing the overdensity profile (dashed lines), suggesting the need for
 fitting (solid lines) or smoothing.}
\label{fig:r500c_est}
\end{figure}

We apply the HSE and corrected mass estimates to a mass-limited sample of 65
galaxy clusters with $M_{500} > 2.2 \times 10^{14}
h^{-1} \msun$ at redshift $z=0$ and their most massive progenitors at $z=0.6$
in the Omega500 simulation\footnote{The sample is chosen with a mass
limit of $M_{\rm 200m} \ge 6\times 10^{14} h^{-1} \msun$ at $z=0$, with $M_{\rm
200m}$ being the mass enclosed in a radius within which the average density is
200 times the mean matter density of the universe.} \citep{nelson14}, an Eulerian
hydrodynamic cosmological simulation. Radiative cooling, star formation and
feedback are not included in the current runs of the Omega500 simulation.
These additional physics are crucial for reproducing the observed
behaviors of cluster cores. They also affect the detailed shapes of the
thermo-dynamical profiles in the bulk of ICM, especially for low mass systems. For low mass clusters
and galaxy groups, testing non-thermal pressure and cluster mass estimations
would be better performed with simulations with cooling and feedback. However,
these additional physics do not change the overall shape of the ICM
profiles outside the cluster core \citep[e.g.][]{nagai07}. In particular, the
velocity dispersion of the gas and fraction of non-thermal pressure at around
$r_{\rm 500}$ are consistent to typically $<$10\% between simulations with and
without cooling and feedback \citep{nagai13,nelson14b} even for
clusters with relatively low masses $M_{\rm 200} \approx 2\times10^{14}$\msun
(corresponding to $M_{\rm 500}\approx
10^{14}h^{-1}\msun$) \citep{bat12}. Therefore, non-radiative simulations are adequate for our
purpose. We refer to Sect.\;2 of SKNN15 and references therein for further
descriptions of the simulation and the cluster sample used in this paper. Small clumps have been removed when constructing the one-dimensional
cluster profiles by excluding high-density tail in the probability distribution
of gas densities using the \citet{zhu13b} method.

We first compute the overdensity profile $\mathbf{\Delta}(r)$ from
simulated profiles with equation (\ref{eq:deltaprof}), taking $P =P_{\rm
th}$ for $M_{\rm HSE}$ and $P = P_{\rm th}/(1 - f_{\rm nth})$
for $M_{\rm corr}$. We then use equation (\ref{eq:mdelta})
to compute $M_{\rm 500}$. Here, $f_{\rm
nth}$ is the `non-thermal fraction' defined by $f_{\rm nth}\equiv P_{\rm
nth}/(P_{\rm th}+P_{\rm nth})$. We compute $f_{\rm nth}$ of each cluster
using the SK14 model and the mass accretion history of the cluster
measured from the simulation (see SKNN15 for details).

Numerically calculating a cluster mass using
equation~(\ref{eq:deltaprof}) is not trivial, as taking derivative of
pressure and dividing by density would amplify small-scale structures and noise.
If we apply no regulation of the simulated profiles other than a cubic-spline
interpolation when computing the derivative of the pressure profile, the
resulting overdensity profile will be far from smooth (dashed lines in
Fig.\;\ref{fig:r500c_est}), despite that the gas density and pressure profiles
are rather smooth after clumps are removed.
In the previous X-ray studies, the observed gas density and temperature
profiles are often fit to parametric models to reduce this noise
\citep[e.g.][]{poi05,vik06,schimdt07}. Non-parametric smoothing methods
have also been adopted in simulation studies \citep{lau09,nelson14}.


Here we use and compare four approaches: one based on cubic spline
smoothing, and the others based on fitting pressure and density to
different parametrized models. For the parametrized models, we further compare
two fitting ranges.

\subsection[]{Cubic spline smoothing}
\label{sec:smoothing}
First, we compute the overdensity profile (equation \ref{eq:deltaprof})
from simulated gas density and pressure profiles with cubic spline
interpolation applied to pressure when taking a derivative, $\partial
P/\partial r$. We then smooth the computed $\log \mathbf{\Delta}(\log
r)$ profile with a cubic-spline smoothing algorithm described in
\cite{dierckx93}. The algorithm fits the data points with a piece-wise
cubic spline with an increasing number of nodes, until the input
constraint on the value of $\chi^2$ statistic is satisfied. We use the
constraint that $\chi^2$ divided by the number of degrees of freedom is
less than 0.04. This constraint is chosen to minimize the variance of
$M_{\rm HSE}/M_{\rm corr}$ while not biasing the sample mean.

When estimating $M_{\rm 500}$ with the smoothing method, a large
radial range of data around the true $r_{\rm 500}$ is needed to guarantee
that the estimated $\Delta(r)$ profile covers $\Delta=500$. Therefore, we apply
this method only to rather relaxed clusters using data on the radial range of
[0.1, 1.5] $r_{\rm 500}^{\rm true}$.

\subsection[]{Parametrized models}
\label{sec:HSEmethods}
Next, we fit parametrized models to the pressure and gas density
profiles and estimate masses. We fit pressure rather than temperature
because pressure is the quantity that directly enters the HSE and corrected mass
estimation. Also, thermal pressure is the quantity that the thermal
Sunyaev-Zel'dovich (SZ) effect \citep{sunyaev70,sunyaev72} directly probes, so
fitting gas density and pressure may be a better option when applying to combined SZ and X-ray
data.

The data points used in the fit are equally spaced logarithmic bins
with uniform weighting on the radial range of either [0.1, 1] $r_{\rm 500}^{\rm
true}$ or [0.1, 1.5] $r_{\rm 500}^{\rm true}$. These are the typical radial ranges of
the currently available data on well-resolved clusters used for the mass
calibration.

\subsubsection[]{Model `VV'}
We first use the model presented in \citet{vik06}, which is based on an
isothermal $\beta$ model \citep{cav78} with modifications capturing the
cool cores and the observed steepening of the gas density profile at large
radii.
Here we ignore the modification for the cool cores, both because cool
cores are non-existent in our simulated clusters, and that our focus is on the
outer region of the clusters, especially around $r_{\rm 500}$ where we determine
their masses.
The resulting simplified model is:
\eq{
\label{eq:vikrho}
\rho_{\rm gas}(r) \propto \frac{1}{(1 + r^2/r_c^2)^{\frac{3\beta}{2}}}
\frac{1}{(1 + r^{\gamma_d}/r_{d}^{\gamma_d})^{\frac{\epsilon}{2\gamma_d}}} \,,
}
and
\eq{
\label{eq:vikP}
P(r) \propto \rho_{\rm gas}(r) \frac{(r/r_t)^{-a}}{(1 +
r^b/r_t^b)^{c/b}} \,.
}
Following \citet{vik06}, we fix $\gamma_d=3$. After this, this model still has
10 parameters (including 2 for normalization), and is highly nonlinear,
suggesting the possible existence of local minima in the $\chi^2$
function.

For all the parametrized models, we fit the simulated gas density and
pressure profiles to the models by searching for the global minimum of the
$\chi^2$ function in a wide parameter range.
Here, we set $0 \leq \epsilon \leq 5$, $0\leq \beta \leq 10$, $-1\leq a \leq 5$, $0\leq b\leq 10$, $0\leq c\leq 10$,
and $r_c$ and $r_t$ being less than 1 Mpc/h.

\subsubsection[]{Model `VG'}
The Vikhlinin model was designed to fit gas density and temperature of the ICM.
It is then not surprising that the fitting form for the pressure
(equation\;\ref{eq:vikP}) is rather redundant. Here, instead,
we use a generalized Navarro-Frenk-White (GNFW) profile \citep{zhao96} for
pressure, as suggested by \citet{nagai07}:
\eq{
P(r) \propto \frac{1}{r^{\gamma}\bb{1+(r/r_{p})^{1/\alpha}}^{\alpha
(\beta-\gamma)}}\,.
\label{eq:pressure-fit}
}
This simpler form already has enough flexibility to describe the shape of the
observed thermal pressure profiles \citep[e.g.][]{arnaud10}, as well as our simulated $P_{\rm
th}$ and $P_{\rm tot}$. Here we further fix $\gamma=0$ which provides an
adequate fit to all our simulated clusters. For the GNFW
parameters, we search for the best fit values within $0.5\leq \alpha \leq 2$,
$3\leq \beta \leq 8$ and $r_{p}$ between 10 kpc/h and 1 Mpc/h.

For the gas density we still use the Vikhlinin formula
(equation \ref{eq:vikrho}). The result is a 9-parameter model.

\subsubsection[]{Model `NG'}
An ideal parametrized model fits the data with enough flexibility with a minimum
number of fitting parameters. The way to achieve this is to utilize existing
knowledge or well-founded assumptions on the underlying physical system that the
data represent.

Here we try to do so by assuming that the total mass density profile follows an
NFW profile \citep{nfw96}. The gravitational acceleration, $g(r)$, is then given
by \eq{
g(r) = \frac{G M(<r)}{r^2} \propto \frac{(1+r/r_s)\ln(1+r/r_s) - r/r_s}{r^2
(1+r/r_s)} \,,
}
where $r_s$ is the scale radius of NFW profile, whose value we search for
between 10 kpc/h and 1 Mpc/h.

For $P_{\rm th}$ and $P_{\rm tot}$ we continue to use GNFW (equation
\ref{eq:pressure-fit}).
The gas density then follows from
$P(r)$ and $g(r)$ as
\eqs{
& \rho_{\rm gas}(r) = -\frac{1}{g(r)}\frac{\dd P(r)}{\dd r} \\
 \propto  & \frac{(1+r/r_s) \bb{\gamma+\beta (r/r_{p})^{1/\alpha}}}
 {\bb{(1+r/r_s)\ln(1+r/r_s) - r/r_s} r^{\gamma-1}
 \bb{1+(r/r_{p})^{1/\alpha}}^{\alpha(\beta-\gamma)+1}} \,.
\label{eq:density-fit}
}
After setting $\gamma=0$, the model has 6 parameters.

The overdensity profiles constructed with this model for one
representative cluster in the sample are shown as the colored solid lines in
Fig.\;\ref{fig:r500c_est}. For this cluster, the fits to the gas density
and {\it total} pressure profiles (i.e., corrected for non-thermal
pressure) give an overdensity profile (blue solid line) that agrees very
well with the true one (black solid line). This suggests that at least for
this cluster, the NFW assumption works and this model provides a good fit to the
simulated profile.

\subsection[]{Dynamical state}
\label{sec:dynstate}

\begin{figure}
\centering
  \begin{tabular}{@{}c@{}}
    \includegraphics[width=0.23\textwidth]{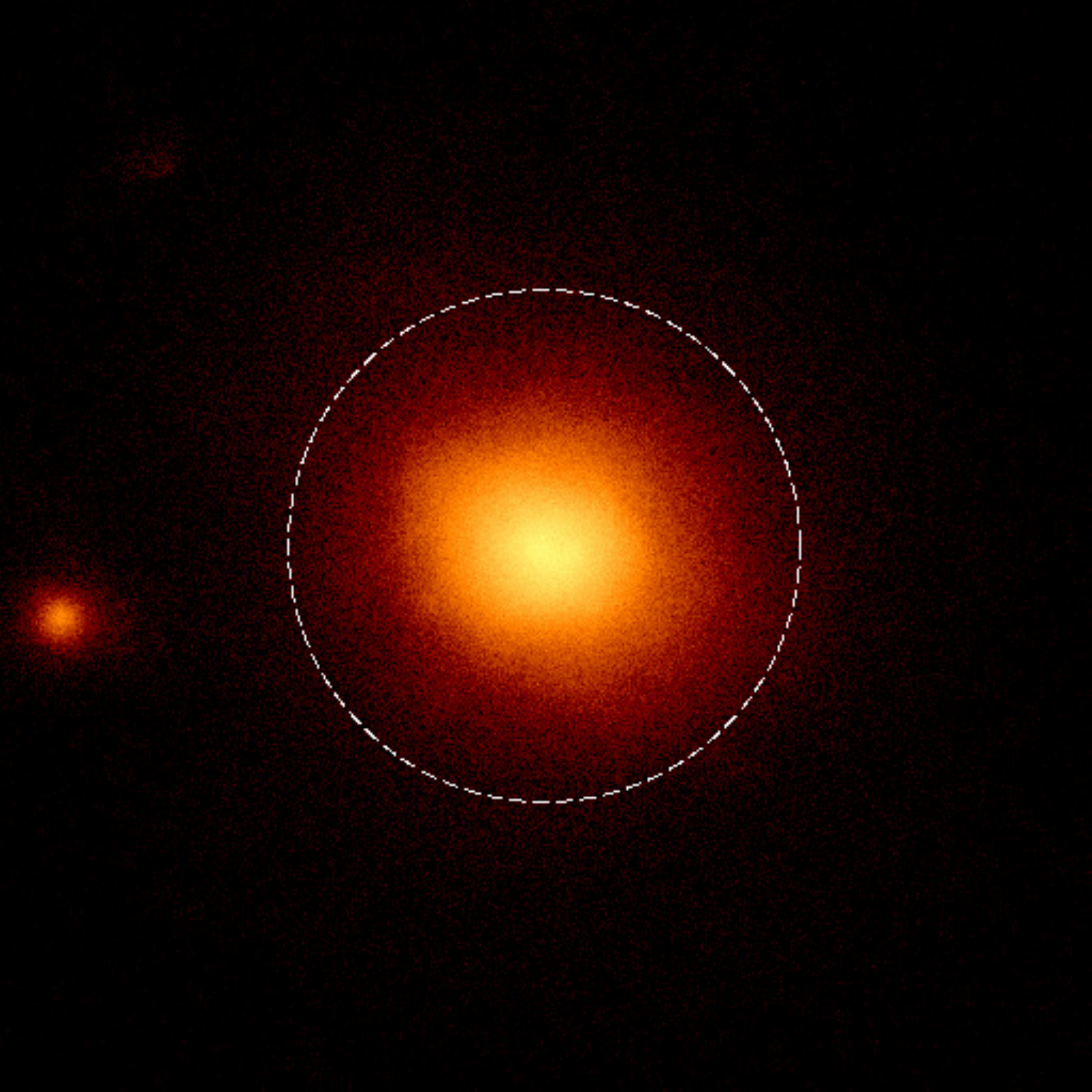}
    \includegraphics[width=0.23\textwidth]{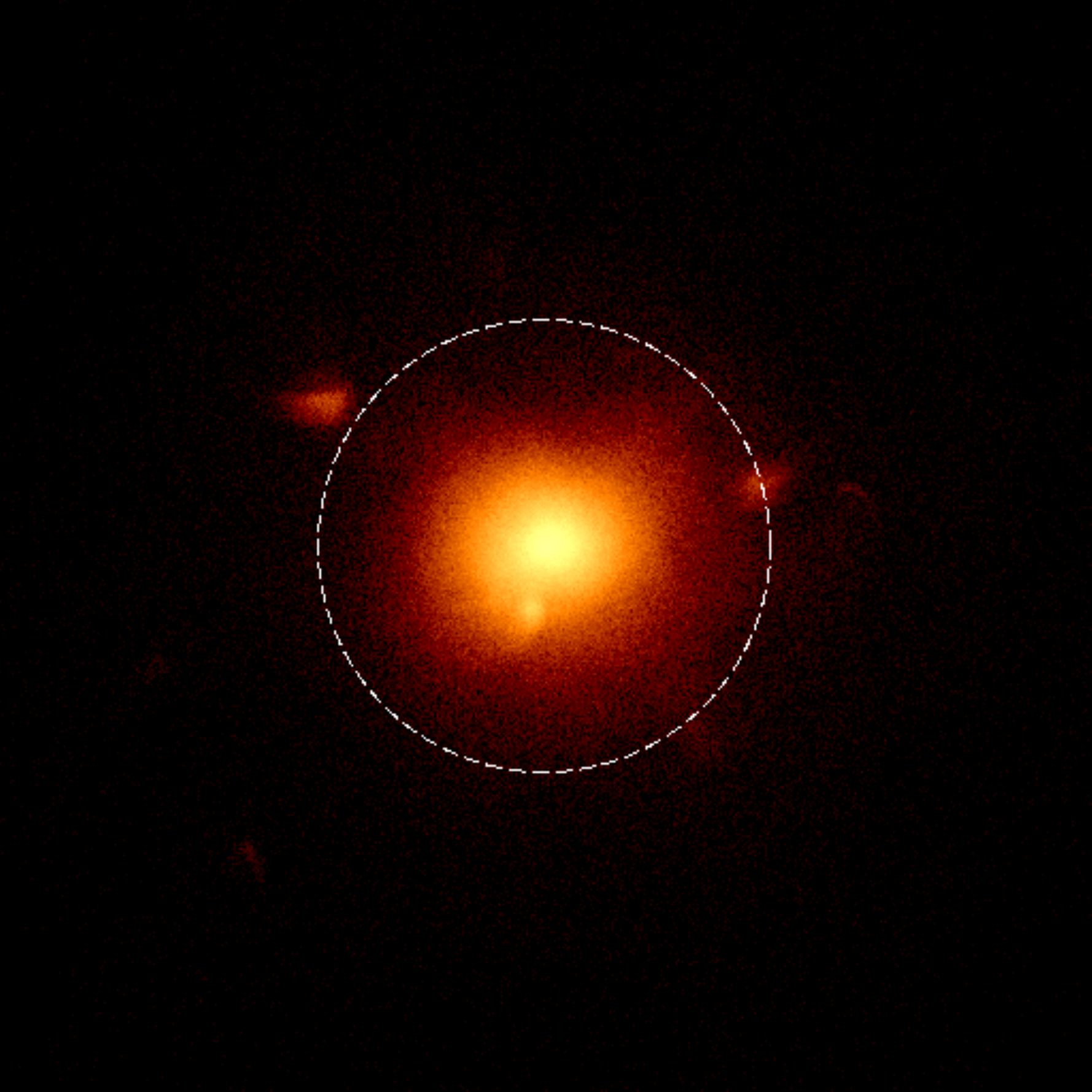}\\
    \includegraphics[width=0.23\textwidth]{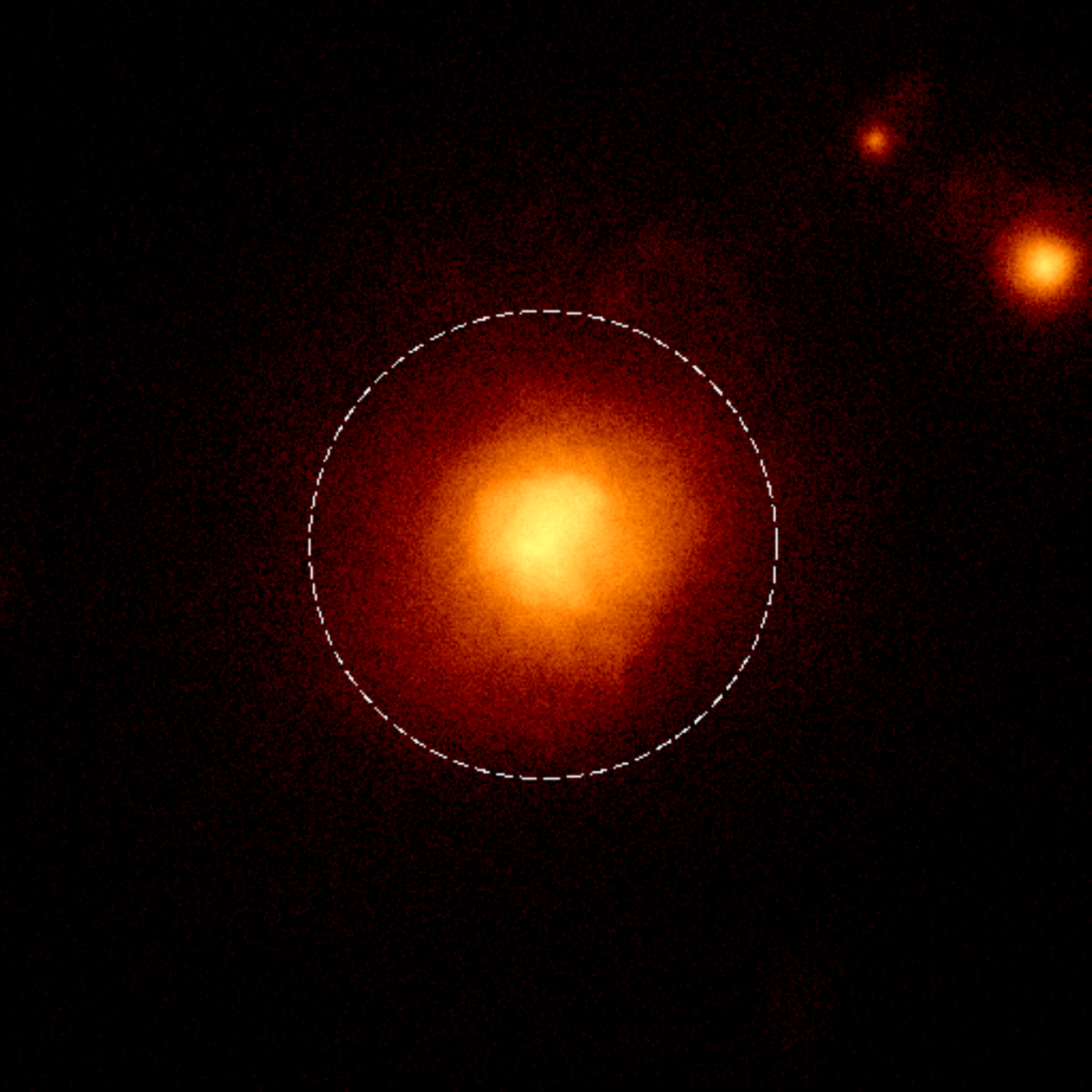}
    \includegraphics[width=0.23\textwidth]{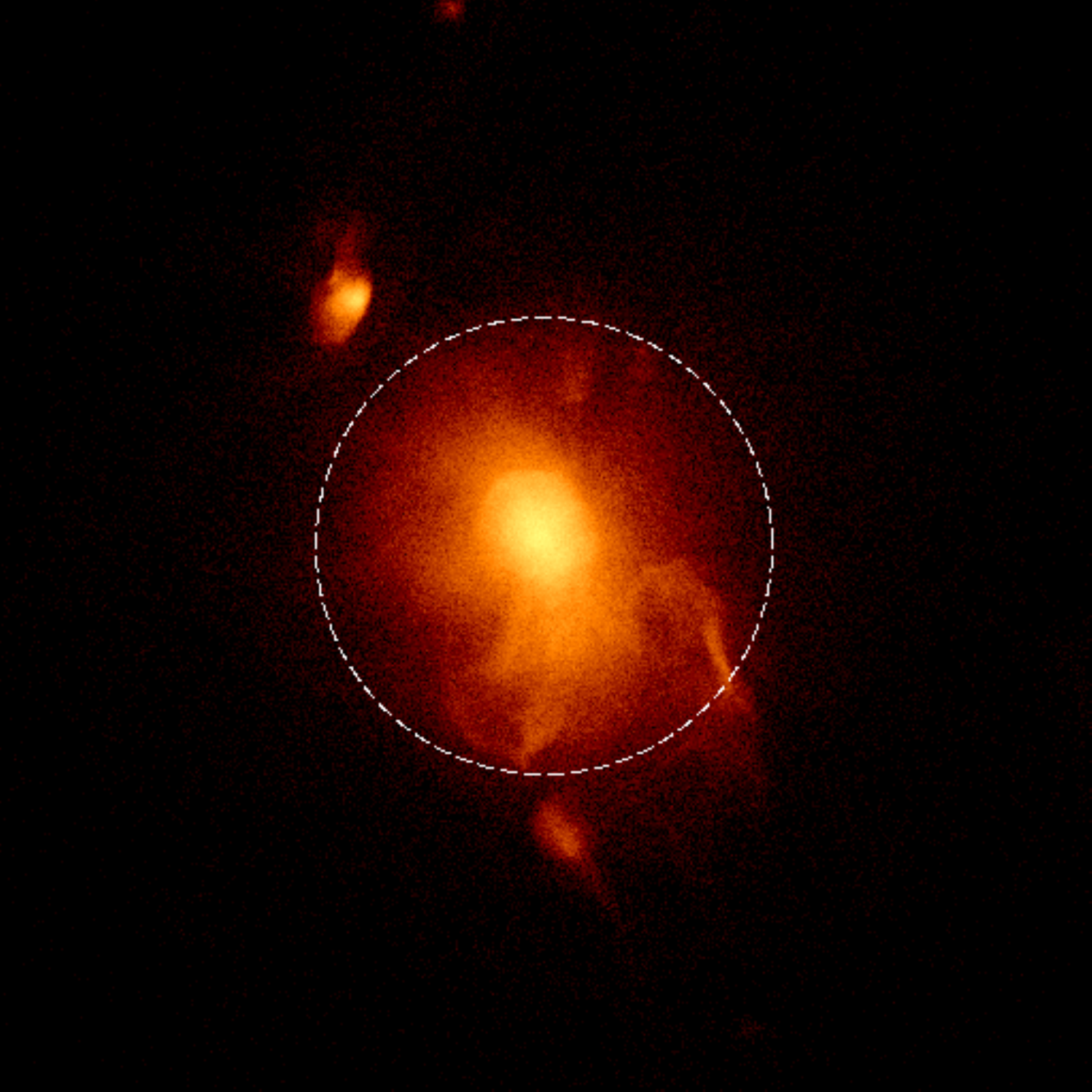}
  \end{tabular}
\caption{Mock X-ray images of four clusters representative of different
dynamical states: `very relaxed' (upper left, cluster number 1 in
Fig.\;\ref{fig:HSEmethod}), `rather relaxed' (upper right, cluster number 5 in Fig.\;\ref{fig:HSEmethod}), `less
disturbed' (lower left), and `disturbed' (lower right). The white dashed
circles show the positions of $r_{\rm 500}$. The exposure time used for the mocks is 100 kilo-seconds.}
\label{fig:dyn}
\end{figure}

Sourced by the growth of cluster via mergers and accretion,
non-thermal pressure correlates strongly with the mass growth history.
The latter also correlates with the dynamical state of the
cluster, but the degree of correlation depends on the particular classification
of dynamical states. Here, we try to mimic the observational procedure and
determine the dynamical state of the clusters based on their X-ray morphologies.
This allows us to estimate the level of non-thermal pressure for the `relaxed'
clusters that are used to calibrate the mass-observable scaling relations. We
compare our classification to that using an automatic method developed by
\citet{mantz15b} in Appendix\;\ref{app:dyn_state}.

We use mock Chandra X-ray images in a single projection, which are produced
using the pipeline developed in \citet{nagai07b}. In particular, we examine (1)
whether the X-ray image shows a single, distinguished cluster core with little displacement with respect to the bulk of the ICM;
(2) whether the cluster core region appears relaxed with round or elliptical
contours; (3) whether there are substructures or clear disturbances in the ICM
between cluster core region and $r_{\rm 500}$; and (4) whether there are
substructures or clear disturbances between [1, 1.5] $r_{\rm 500}$ that
will influence the pressure gradient estimation at $r_{\rm 500}$.

Only 1 out of the 65 clusters appears very relaxed according to all the above
criteria. If we relax the criteria to allow for the existence of small
concentrated substructures that can be removed by the
clump-removing method, then 3 more clusters can be classified as `very relaxed'.
We construct larger sub-samples of clusters by further relaxing the criteria: a
`rather relaxed' sample of 14 clusters which have no or only slight
disturbances in the core region and the global ICM, and contain no or only
small concentrated substructures within and around $r_{\rm 500}$; and a `less
disturbed' sample of 32 clusters (including the `rather relaxed' clusters)
which have single, distinguished cluster core and no significant substructure
between cluster core region and $r_{\rm 500}$. Fig.\;\ref{fig:dyn} shows
 mock Chandra X-ray images of representative clusters of these samples.

\subsection[]{Results}
\label{sec:results}
\begin{figure}
\centering
  \begin{tabular}{@{}c@{}}
    \includegraphics[width=.45\textwidth]{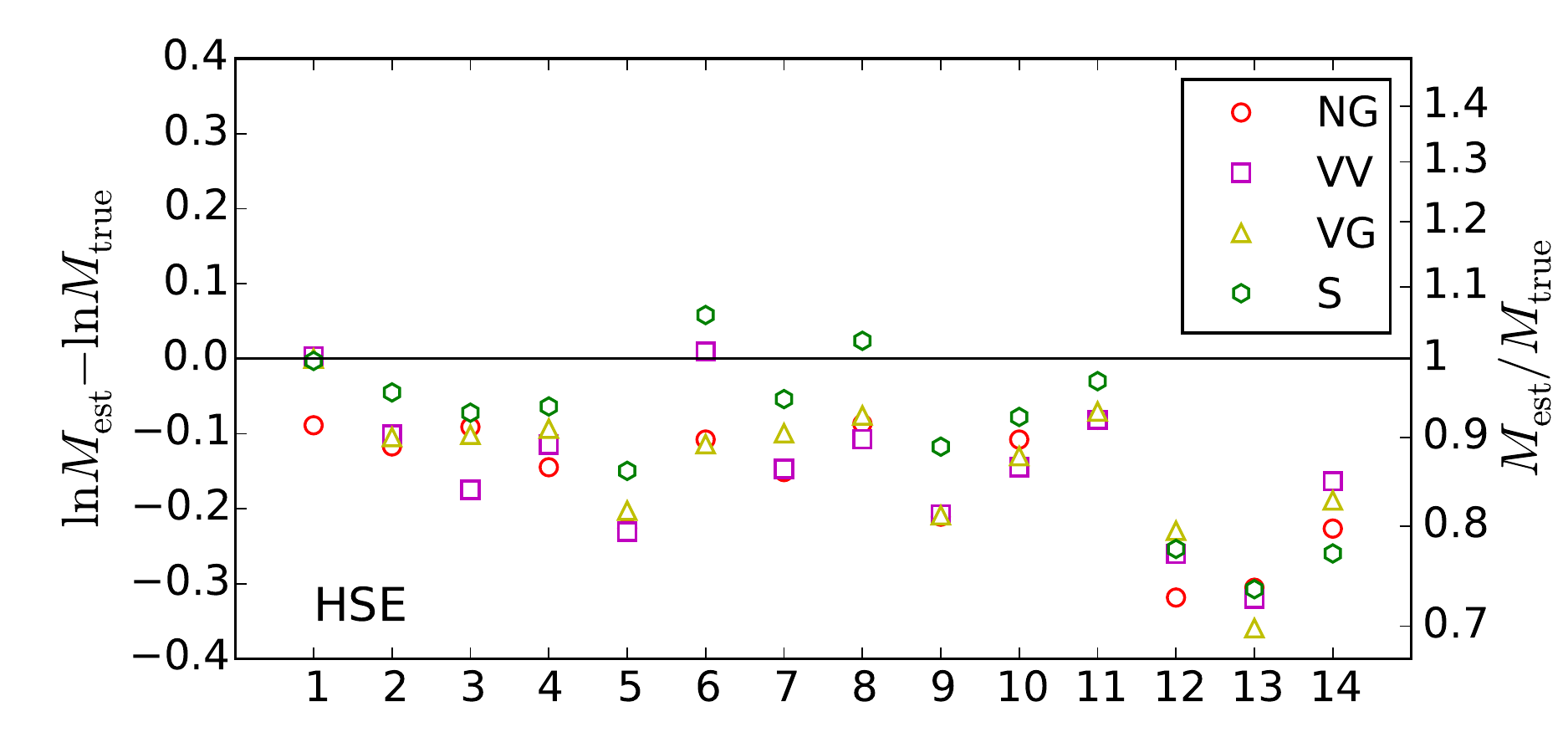}   \\
    \includegraphics[width=.45\textwidth]{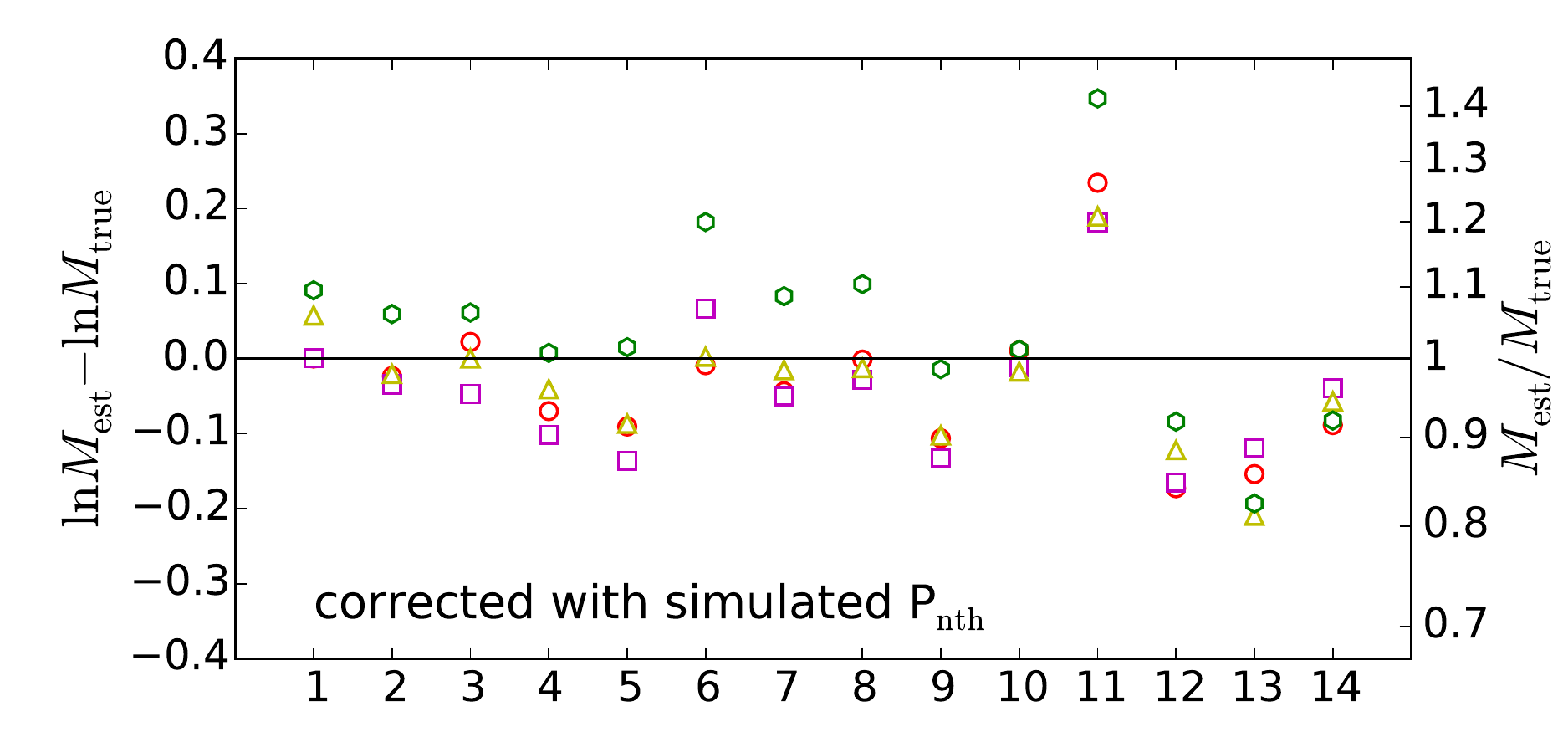}   \\
    \includegraphics[width=.45\textwidth]{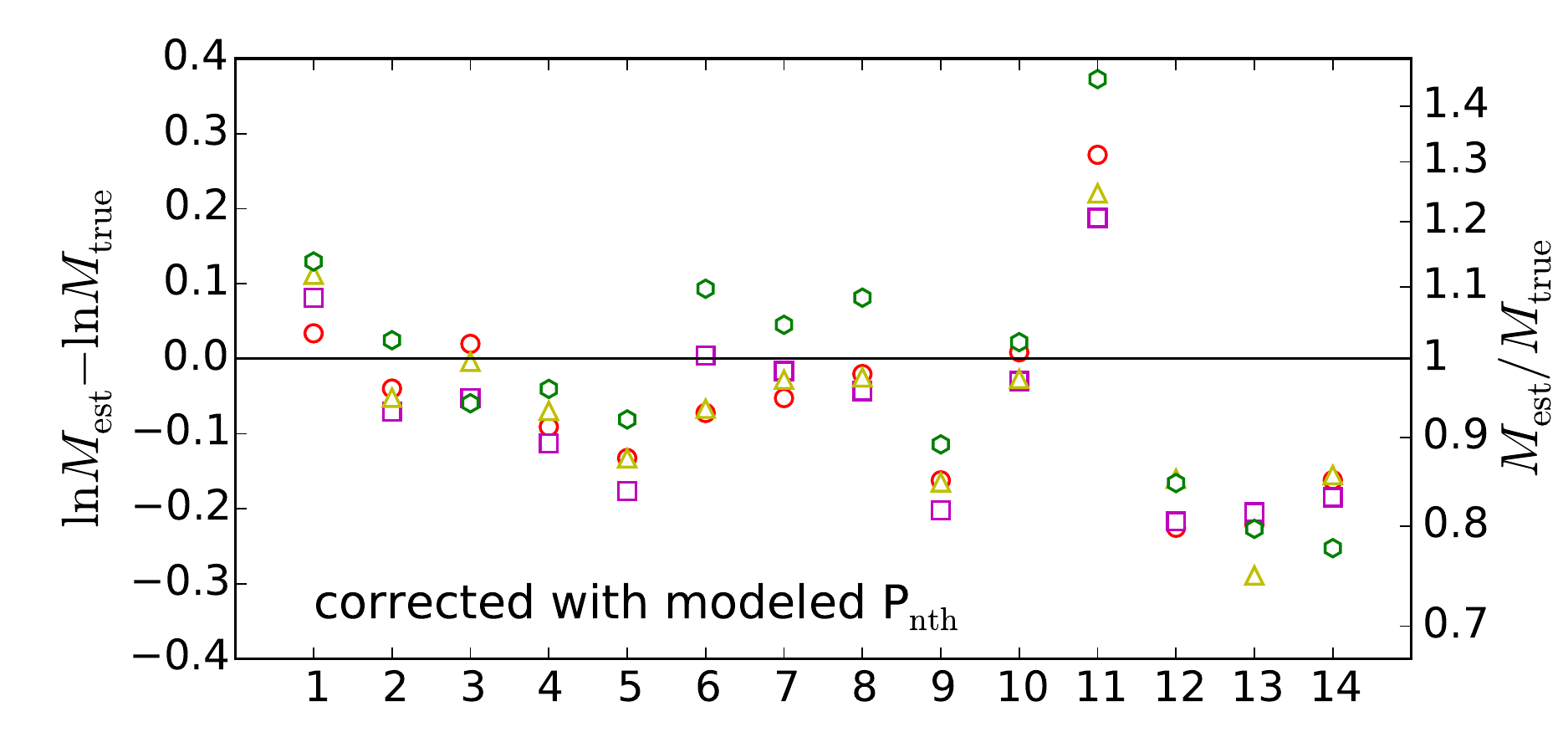}
  \end{tabular}
  \caption{HSE mass biases at $r_{\rm 500}$ (upper panel) and their correction
  with the simulated non-thermal pressure (middle panel) and the modeled non-thermal pressure
  (lower panel) for the `rather relaxed' clusters.
  The 14 clusters are ordered by their degree of relaxation, ranging from very
  relaxed on the left (number 1) to slightly disturbed on the right (number 14). The masses are estimated using profiles
  between 0.1 and 1.5 $r_{\rm 500}^{\rm true}$ with four methods mentioned in
  the text:
  a spline smoothing method (S) and three parametrized model fitting methods (NG,
  VV and VG). The averaged mass biases for the sample before and after
  correction are given in Fig.\;\ref{fig:biasall} and
  Table\;1.}
\label{fig:HSEmethod}
\end{figure}

\begin{figure}
\centering
    \includegraphics[width=.45\textwidth]{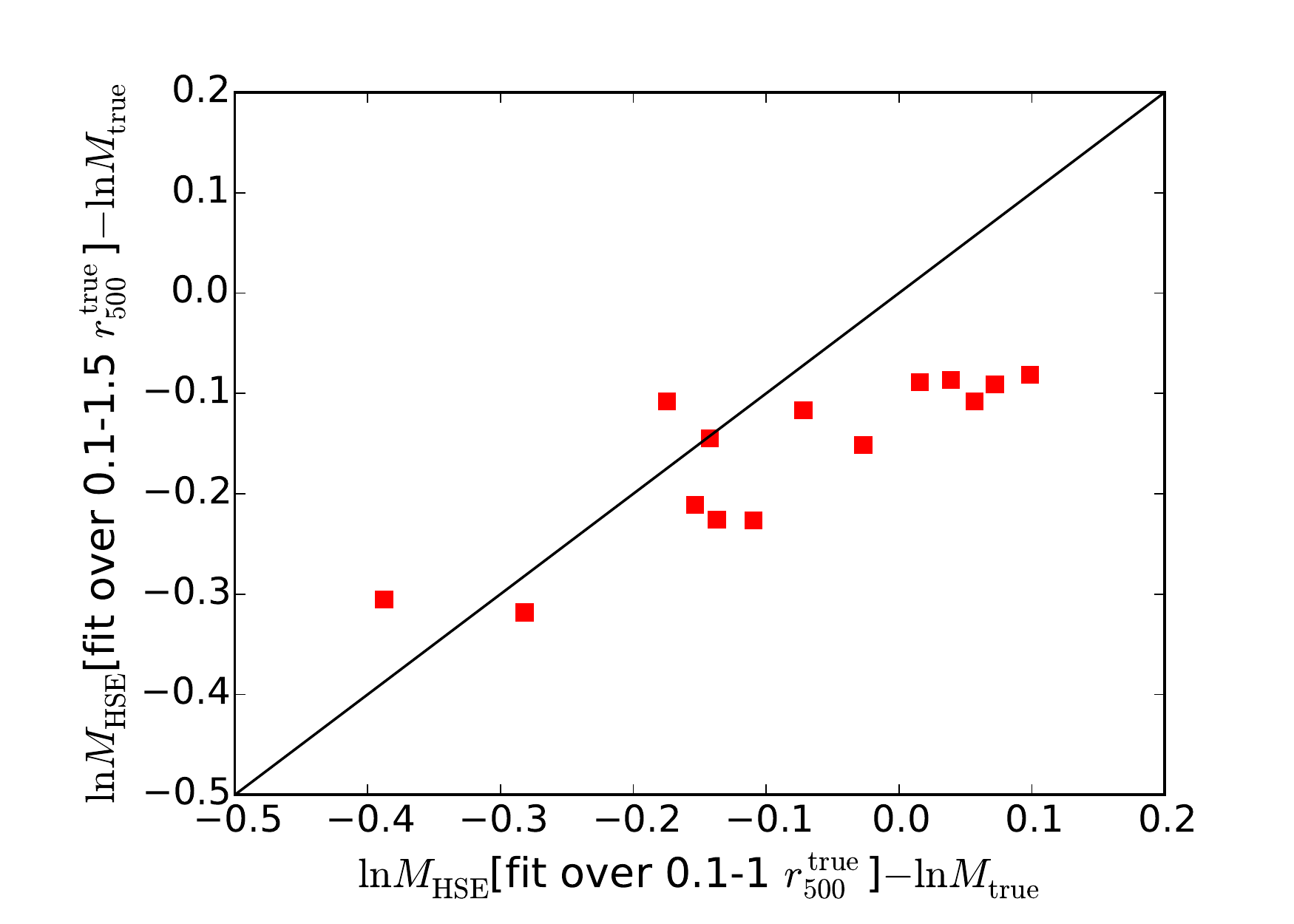}
  \caption{Effect of fitting range on the HSE mass bias.
  HSE mass biases at $r_{\rm 500}$ of the `rather relaxed' 14 clusters estimated
  using the parametrized fitting method `NG' are shown.}
\label{fig:HSEradial}
\end{figure}

\begin{figure}
\centering
    \includegraphics[width=.49\textwidth]{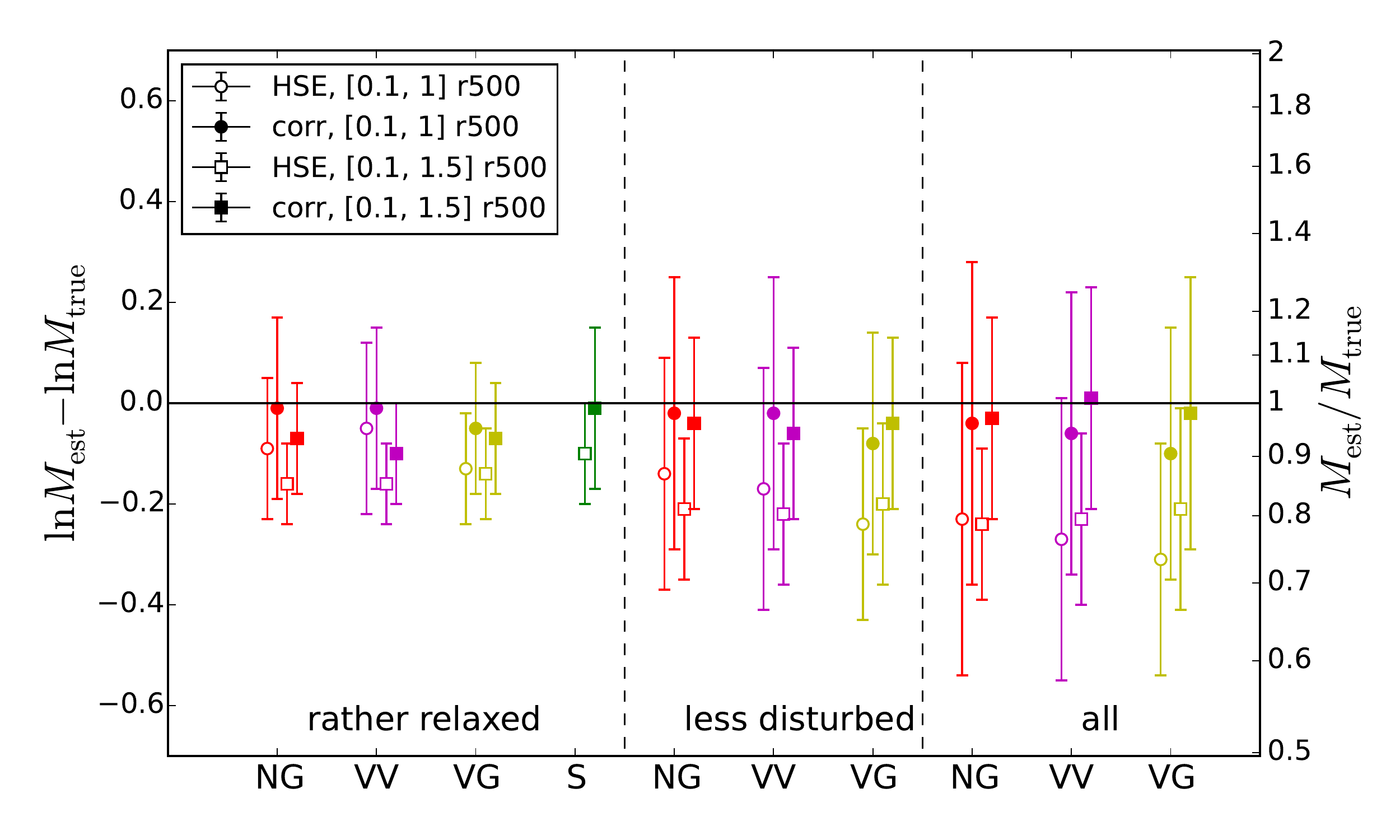}
  \caption{Averaged HSE (open markers) and corrected (filled markers) mass
  biases at $r_{\rm 500}$ and their scatter for cluster samples of different
  relaxation states:
  top 14 (`rather relaxed') and 32 (`less disturbed'), and the full mass limited
  65 clusters (`all'). The fitting / smoothing method is indicated by the label
  on the x-axis, with symbols identical to that in Fig.\;\ref{fig:HSEmethod}.
  Fitting methods are performed using profiles on two different radial ranges: [0.1, 1] $r_{\rm 500}^{\rm true}$ (circles) and [0.1, 1.5] $r_{\rm
  500}^{\rm true}$ (squares). The numerical values are given in Table\;1.}
\label{fig:biasall}
\end{figure}

\begin{table*}
 \centering
{
\caption{Numerical values of Bias~$\pm$~Scatter in mass estimations with the
fitting method `NG' shown by the red symbols and error bars in
Fig.\;\ref{fig:biasall}. The values for the progenitors at $z=0.6$ are also
shown in the last column.}
\begin{tabular}{cccccc}
 \hline

  & fitting range  & rather relaxed (14/65) &  less disturbed (32/65) &
  mass-limited (65/65) & progenitors at z=0.6 (65/65) \\
\hline

$\ln(\MHSE/\Mtrue)$  & [0.1, 1] $r_{\rm{500}}^{\rm true}$	& $-0.09\pm0.14$	&
$-0.14\pm0.23$	& $-0.23\pm0.31$ &	 $-0.24\pm0.26$ \\

					 & [0.1, 1.5] $r_{\rm{500}}^{\rm true}$	& $-0.16\pm0.08$	& $-0.21\pm0.14$
					 & $-0.24\pm0.15$  & $-0.26\pm0.15$ \\
\hline
$\ln(\Mcorr/\Mtrue)$ & [0.1, 1] $r_{\rm{500}}^{\rm true}$	& $-0.01\pm0.18$
& $-0.02\pm0.27$	&$-0.04\pm0.32$ &  	$-0.07\pm0.28$ \\

					 & [0.1, 1.5] $r_{\rm{500}}^{\rm true}$	& $-0.07\pm0.11$	& $-0.04\pm0.17$
					 &$-0.03\pm0.20$  &$-0.05\pm0.18$ 	\\
\hline
\end{tabular}}
\label{tab:mbias}
\end{table*}

\subsubsection[]{Influence of mass estimation methods on the HSE mass bias}
In the upper panel of Fig.\;\ref{fig:HSEmethod} we show the HSE mass biases of
the 14 `rather relaxed' clusters. Given the same input gas density and pressure
profiles, the HSE masses estimated using different approaches
(Sect.\;\ref{sec:smoothing} and \ref{sec:HSEmethods}) vary at the level of
5-10\% for all 14 clusters. None of the four approaches systematically give
significantly smaller or larger HSE mass biases, suggesting that they are
comparable in a statistical sense. Even for the few most relaxed
clusters (cluster number 1-4), there exist 5-10\% HSE mass biases, confirming
the need for non-thermal pressure correction even for the relaxed cluster
sample typically used for calibrating the scaling relation.

Fitting to different radial ranges
also introduces 5-10\% differences in the HSE mass estimates
(Fig.\;\ref{fig:HSEradial}). Fitting to larger radii more likely gives
larger HSE mass biases. This is because the increase of non-thermal pressure
with radius causes additional steepening of $P_{\rm th}$ compared to $P_{\rm tot}$, so when
data from larger radii are considered in the fit, larger HSE mass biases are
favored given the limited flexibility of any fitting formula. Being
a weak effect, this systematic trend is visible only for relatively relaxed clusters
(Fig.\;\ref{fig:biasall}).

\subsubsection[]{Correction for the HSE mass bias}
How well can we correct for the HSE mass bias with the SK14 analytical model for
non-thermal pressure? The lower panel of Fig.\;\ref{fig:HSEmethod} shows
the corrected masses for the `rather relaxed' clusters. The correction for individual clusters is not
perfect, which is also the case if we correct for the masses using the
{\it simulated} $P_{\rm nth}$ (middle panel). The imperfection of HSE mass
correction with simulated $P_{\rm nth}$ is also noticed in \citet{nelson14}.
This imperfection is most likely due to asphericity and the small scale
structure of the ICM as well as residual accelerations which also breaks the HSE
assumption \citep{lau13}.
However, the average HSE mass bias for the 14 clusters is greatly reduced after
correction, both with simulated and modeled $P_{\rm nth}$, irrespective of
which smoothing/fitting method is used.
This is more apparent from
Fig.\;\ref{fig:biasall}, where we show the averaged HSE (open markers) and
corrected (filled markers) mass biases and their scatter. Remarkably, for the
`less disturbed' and even the full mass-limited sample, both at z=0 (5th column
of Table\;1) and a high redshift of z=0.6 (6th column), the biases are successfully reduced
to be well below 10\% irrespective of the smoothing/fitting method or the fitting range
(Fig.\;\ref{fig:biasall}; Table\;1). This suggests that, the HSE mass biases related to non-thermal
pressure can be corrected robustly for the sample average with the SK14
model for clusters in a wide range of dynamical states and redshifts, which will
be obtained with combined X-ray surface brightness and SZ observations.

\section{Conclusions and Discussions}
\label{sec:conclusion}

There is no universal value for the HSE mass bias due to non-thermal
pressure. Not only does it depend on masses, radii, and redshifts (SK14),
but it also depends on the particular fitting or smoothing method used
to compute the HSE mass from the observed profiles, as well as on the radial
range of the data used for the fitting. Thus, to evaluate how much non-thermal
pressure contributes to the `Planck CMB-cluster tension', the
particular method used by the Planck team needs to be tested e.g. on simulated
cluster samples.

Non-thermal pressure biases the masses, hence the mass-observable scaling
relations, even for the most relaxed clusters in the mass-limited sample in the
simulation.
This poses a challenge to the commonly adopted approach of calibrating the
scaling relation on a sample of relaxed clusters and applying it to the full
cluster sample at all dynamical states. Correction for non-thermal pressure is
needed for the calibration of the scaling relation on relaxed clusters
too because non-thermal motions are continuously sourced by the mass
accretion and are never fully thermalized, especially at the cluster outskirts.

The SK14 model is able to correct the non-thermal pressure biases in the
mass-limited sample and its dynamically more relaxed subsamples, irrespective of
the dynamical state, the smoothing or the fitting method used for the mass
estimation, or the radial range of data used in the fitting.
Aided by this correction, the HSE mass estimate may no longer be limited by
non-thermal pressure or dynamical state of the cluster. This opens up a
possibility for more accurate cluster mass determinations using combined X-ray surface
brightness and SZ observations. Compared to cluster masses
estimated using weak gravitational lensing, the bias-corrected HSE masses are
more precise, i.e. have smaller statistical uncertainties. Nevertheless, it
will be important to check the consistency of both methods.

Gas motions in the ICM will be probed by the upcoming ASTRO-H mission
\citep{tak12} through the shifts and broadening of the X-ray spectral lines \citep{zu15,ota15}.
However, the measurements will be limited to a small number of clusters and small radii ($r\lesssim r_{\rm
2500}$) where the random motions are mild. Although the measurements cannot be
directly used to correct mass estimations at $r_{\rm 500}$, they will test
our current understanding on the nature of gas motions in the ICM.

The necessary input of the SK14 model is the mass accretion
history of clusters. Although not directly observable, the mass accretion
history is known in a sample-averaged sense from dark matter-only cosmological
simulations \citep[e.g.][]{zhao09}, though its cosmology dependence awaits more
detailed simulation studies.
Our simple correction method reduces the mean
of the mass biases but not yet the scatter, which means no advantage will be
gained by correcting HSE mass biases for individual clusters at this stage. If the
residual acceleration, asphericity and small scale structures of the ICM, which
are the most likely sources of the scatter, are better understood in the
future, then knowledge of non-thermal pressure in individual clusters will be very
helpful in getting accurate individual masses.
The mass accretion history and the non-thermal pressure for individual
clusters may also be obtainable by exploiting the connection between the mass density profile
and the mass accretion history \citep{lud13,diemer14}. Namely, first compute a
mass profile from a cluster without correcting for non-thermal pressure, and
then compute the mass accretion history from the mass profile, compute
non-thermal pressure, and recompute the mass profile. Then iterate until the
mass profile converges. We leave this for future work.

\section*{Acknowledgements}
We thank Kaylea Nelson for providing the simulation data used
in this work. DN thanks Max-Planck-Institut f\"ur Astrophysik for hospitality when
this work is finished. This work was supported in part by NSF grant AST-1412768,
the Research Corporation, and by the facilities and staff of the Yale Center for Research Computing.
\bibliographystyle{mn2e}

\appendix
\section{Velocity anisotropy}
\label{sec:anisotropy}

In the absence of streaming motions, the steady-state radial Jeans equation
gives \citep[see e.g.][]{lau13}

\eq{
\label{eq:Mrand}
\frac{G M(<r)}{r^2}  =  - \frac{1}{\rho_{\rm gas}} \frac{\partial
\br{P_{\rm th} + \rho_{\rm gas} \sigma^2_r }}{\partial r} - \frac{\br{2
\sigma^2_{r} - \sigma^2_{t} }}{r} \,,
}
where $\sigma^2_{r}$ is the radial non-thermal velocity dispersion, and
$\sigma^2_{t} = \sigma^2_{\theta \theta} + \sigma^2_{\phi\phi}$ is the
tangential non-thermal velocity dispersion.

Let us define the velocity anisotropy parameter, $\beta$, as
\eq{
\label{eq:beta}
\beta \equiv 1 - \frac{\sigma^2_{t}}{2\sigma^2_{r}}\,.
}
This parameter vanishes for isotropic velocity distribution, and equation~(\ref{eq:Mrand}) reduces to equation~(\ref{eq:Mcorr}) with $P_{\rm nth} = \rho_{\rm
gas} \sigma_{\rm nth}^2$. Here, $\sigma^2_{\rm nth} \equiv (\sigma^2_{r} +
\sigma^2_{t})/3$ is the averaged one-dimensional non-thermal velocity
dispersion squared.

We calculate the influence of a non-zero anisotropy parameter on the HSE mass
estimate.
Inserting equation~(\ref{eq:beta}) into equation~(\ref{eq:Mrand}), we obtain
\eqs{
& M(<r) = -\frac{r^2}{G \rho_{\rm gas}}\frac{\partial P_{\rm
th}}{\partial r} -\frac{3}{3-2\beta} \frac{r \sigma^2_{\rm nth}}{G} \times
\\
&  \bb{\frac{\partial\ln\rho_{\rm gas}}{\partial\ln r} +
\frac{\partial\ln\sigma^2_{\rm nth}}{\partial\ln r} -
\frac{\partial\ln\br{3-2\beta}}{\partial\ln r} + 2\beta} \,.
}
When $\beta\neq 0$, the difference between the mass estimates without
assuming $\beta=0$, $M_\beta$, and assuming $\beta=0$, $M_{\beta=0}$, is
given by
\eqs{
&-\frac{G}{r \sigma^2_{\rm nth}} \left[{M_{\beta}(<r) - M_{\beta=0}(<r)}\right] =
  \frac{2\beta}{3-2\beta} \frac{\partial\ln\rho_{\rm gas}}{\partial\ln r} +
\frac{6\beta}{3-2\beta}  \\
&
+ \frac{2\beta}{3-2\beta} \frac{\partial\ln\sigma^2_{\rm nth}}{\partial\ln r}
 - \frac{3}{3-2\beta} \frac{\partial\ln\br{3-2\beta}}{\partial\ln r} \,.
}
The first two terms on the r.h.s. dominate; however, they cancel
when $\rho_{\rm gas}$ has a logarithmic slope of $-3$, which
happens to coincide with the outer slope of an NFW profile. At $r_{\rm 500}$,
$\rho_{\rm gas}$ typically has a slightly shallower logarithmic slope than $-3$,
but still letting these terms cancel to a large extent.
Therefore, the difference in mass corrections with and without including
velocity anisotropy is expected to be smaller than the mass
correction itself, which is
\eq{
-\frac{G}{r \sigma^2_{\rm nth}} \bb{M_{\rm corr}(<r) - M_{\rm HSE}(<r)} =
 \frac{\partial\ln\rho_{\rm gas}}{\partial\ln r} +
\frac{\partial\ln\sigma^2_{\rm nth}}{\partial\ln r}
\,.
}

\begin{figure}
\centering
  \begin{tabular}{@{}c@{}}
    \includegraphics[width=.47\textwidth]{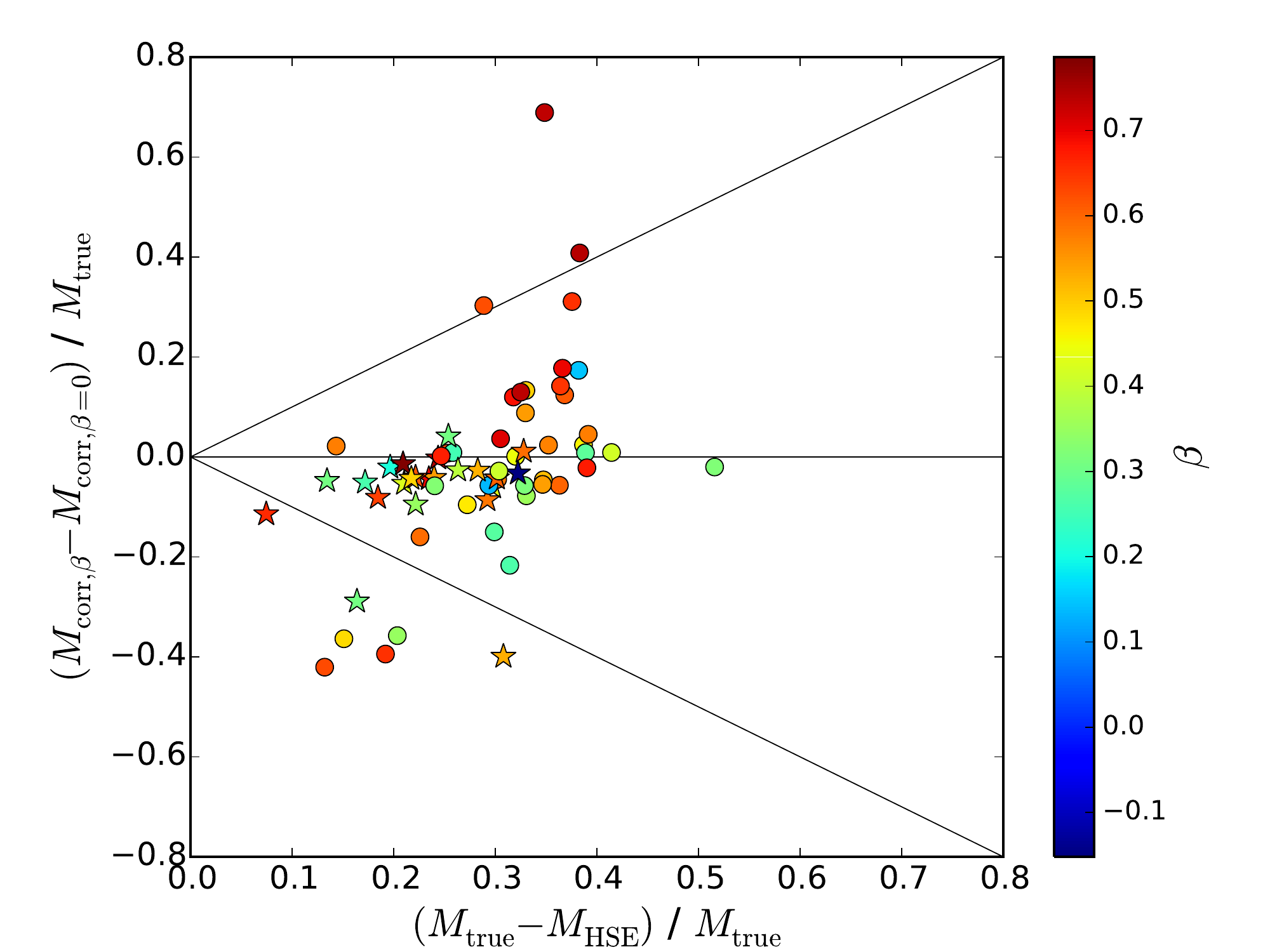}
  \end{tabular}
  \caption{Difference in the $M_{\rm corr}$ obtained by assuming and not
  assuming $\beta=0$ when correcting for non-thermal pressure is smaller than the correction
 itself, $M_{\rm true}-M_{\rm HSE}$, for the majority of clusters in the
 simulation. The color scale shows $\beta$: red
 for bigger $\beta$ and blue for smaller $\beta$. The stars show the
 subsample of relaxed clusters, while the circles show the rest. Only 10 out
 of 65 clusters show $|M_{\rm corr,\beta}-M_{\rm corr,\beta=0}|>M_{\rm
 true}-M_{\rm HSE}$.
}
  \label{fig:beta500c}
\end{figure}

In this Appendix, we use the cubic spline smoothing method to
regularize the simulated cluster profiles in the radial range of [0.01, 3]
$r_{\rm 500}^{\rm true}$.
Fig.~\ref{fig:beta500c} shows that the difference between assuming and not assuming $\beta=0$ is indeed smaller
than the correction itself, i.e., $M_{\rm true}-M_{\rm HSE}$, for the
majority of clusters. Even in the cases where the difference is significant, the
mass estimated by taking into account the velocity anisotropy is {\it not}
always closer to the true mass. Thus including velocity anisotropy does not give significantly
improved results. This, together with that $\beta(r)$ is not observable,
makes us conclude that the simple mass correction given in equation~(\ref{eq:Mcorr}) would already suffice.

\section[]{From non-thermal pressure to mass bias}
\label{app:massbias}

\begin{figure}
\centering
\includegraphics[width=0.48\textwidth]{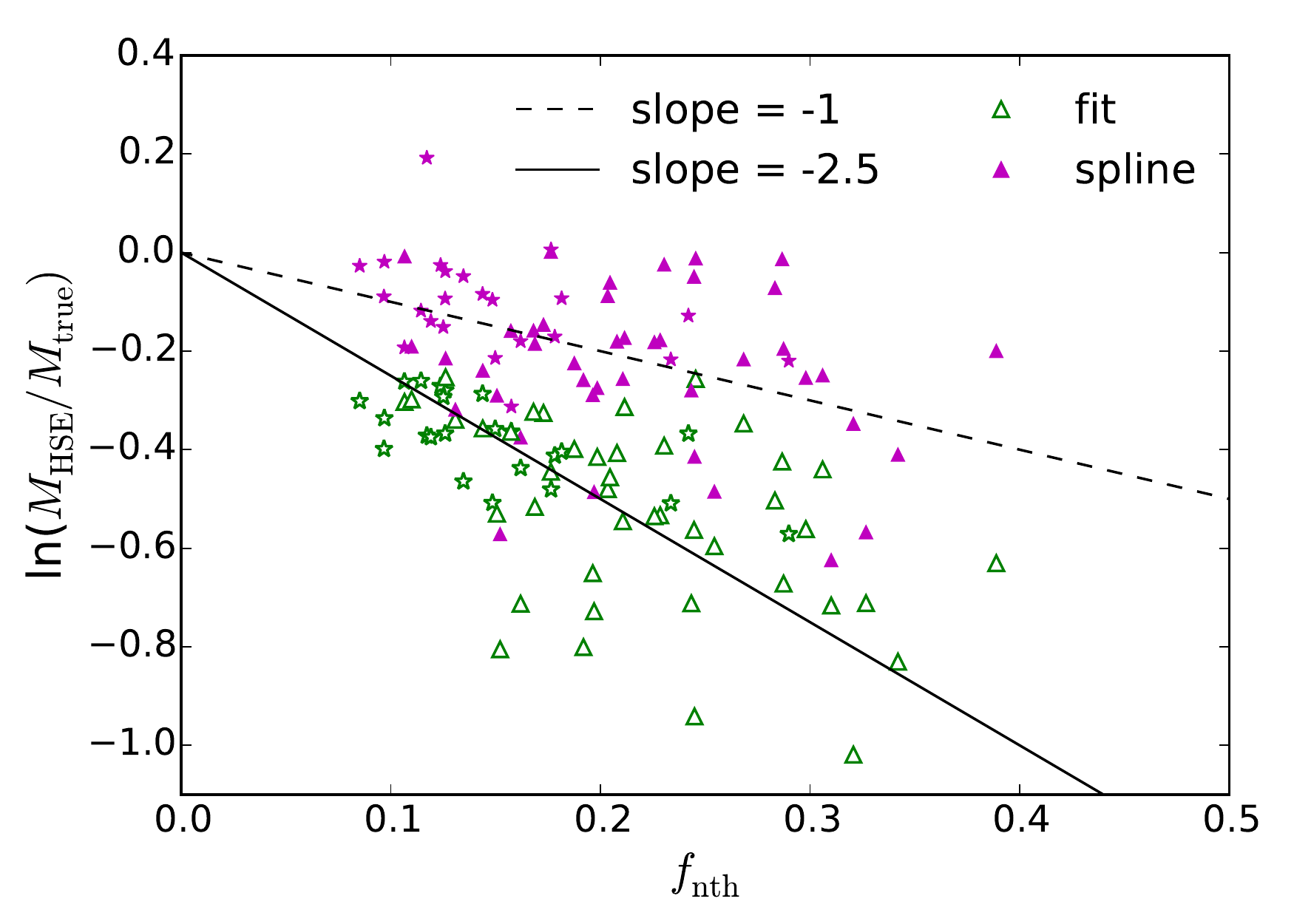}
\caption{Correlation between non-thermal pressure and the HSE mass bias
 at $r_{500}$. The open and filled symbols show $M_{\rm HSE}/M_{\rm
 true}$ from the `NG' parametric fitting  and the cubic spline smoothing
 methods, respectively, as a function of the non-thermal fraction, $f_{\rm nth}$,
 at $r_{500}$. The stars show the subsample of relaxed clusters, while
 the triangles
 show the rest. The dashed and solid lines show the slopes of $-1$ and
$-2.5$ to guide the eye.}
\label{fig:dlnM_fnth}
\end{figure}

In this Appendix, we show how an error on the pressure profile
propagates to an error on the mass. According to
equation~(\ref{eq:deltaprof}), an error on the pressure gradient will
lead to an error on $\mathbf{\Delta}(r)$ as $\delta \ln
\mathbf{\Delta} = - \delta \ln(\partial P/\partial r)$. Since the error on the estimation of $r_{\Delta}$ is related to the error on the
$\mathbf{\Delta}(r)$ profile through the chain rule
$\delta \ln(r) = \delta \ln \mathbf{\Delta}(r)\;  (\partial \ln
\mathbf{\Delta} / \partial\ln r)^{-1}$, the cluster mass will be
miss-estimated by

\eq{
\delta \ln (M_{\Delta}) = 3\delta \ln(r_{\Delta}) = - 3\delta \ln \br{
\frac{\partial P}{\partial r}}  \br{\frac{\partial \ln \mathbf{\Delta}(r)}{\partial\ln r}}^{-1}
\bigg|_{r_{\Delta}} \,.
}
The slope of $\mathbf{\Delta}(r)$ is approximately $-2$ ($\pm 0.2$) at a radius
of $r_{\rm 500}$ for an NFW mass density profile with a concentration parameter
typical of clusters.
Hence, $\delta \ln(M_{\Delta}) \approx 3 \delta
\ln \br{\partial P / \partial r} /2$.

The relation between the change in the pressure gradient $\delta
\ln \br{\partial P / \partial r}$ and the non-thermal fraction $f_{\rm nth}$ is:
\eqs{
\label{eq:slope_fnth}
& \delta \ln \br{ \frac{\partial P}{\partial r}}  \equiv \ln \br{ \frac{\partial
P_{\rm th}}{\partial r}} - \ln \br{ \frac{\partial P_{\rm tot}}{\partial r}}
\\
 = & \ln (1 - f_{\rm nth}) + \ln \bb{ 1 + \br{\frac{\partial \ln P_{\rm
tot} }{\partial \ln r}}^{-1} \frac{\partial \ln (1-f_{\rm nth})}{\partial \ln
r} }\,.
}
In the limit that $f_{\rm nth}$ is small and constant with radius, the
relation is linear: $\delta \ln(M_{\Delta}) \approx -3f_{\rm nth}/2$. In
practice, $f_{\rm nth}$ is radial dependent \citep{shi14}, and this relation
also depends on the method used in regularizing the input cluster profiles. We
show in Fig.\;\ref{fig:dlnM_fnth} that $|\delta \ln (M_{\Delta})|$ estimated with both
the `NG' fitting and the cubic spline smoothing methods correlate positively
with $f_{\rm nth}$, and the fitting method yields a significantly larger HSE
mass bias since we fit out to a large radius of 3$r_{\rm 500}^{\rm true}$ here.

If the HSE mass estimation is performed at the true $r_{\Delta}$  (as e.g. in
Sect.\;6 of SK14), the mass bias is $\delta \ln(M_{\Delta}) =  \delta
\ln \br{\partial P / \partial r}$, i.e., about 2/3 of the above value.

\section[]{Comparison of dynamical state classifications}
\label{app:dyn_state}
\begin{figure}
\centering
  \begin{tabular}{@{}c@{}}
    \includegraphics[width=0.33\textwidth]{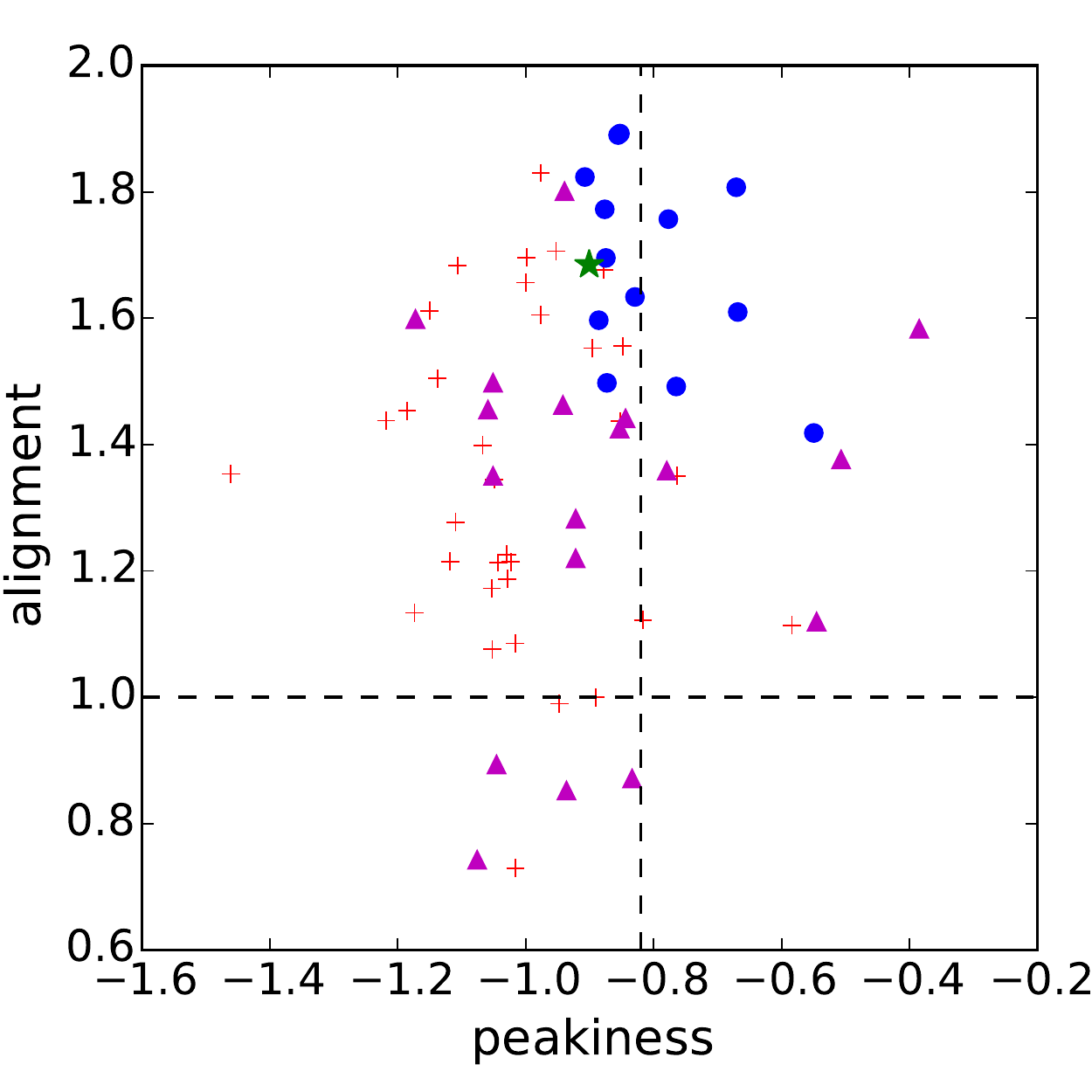}\\
    \includegraphics[width=0.33\textwidth]{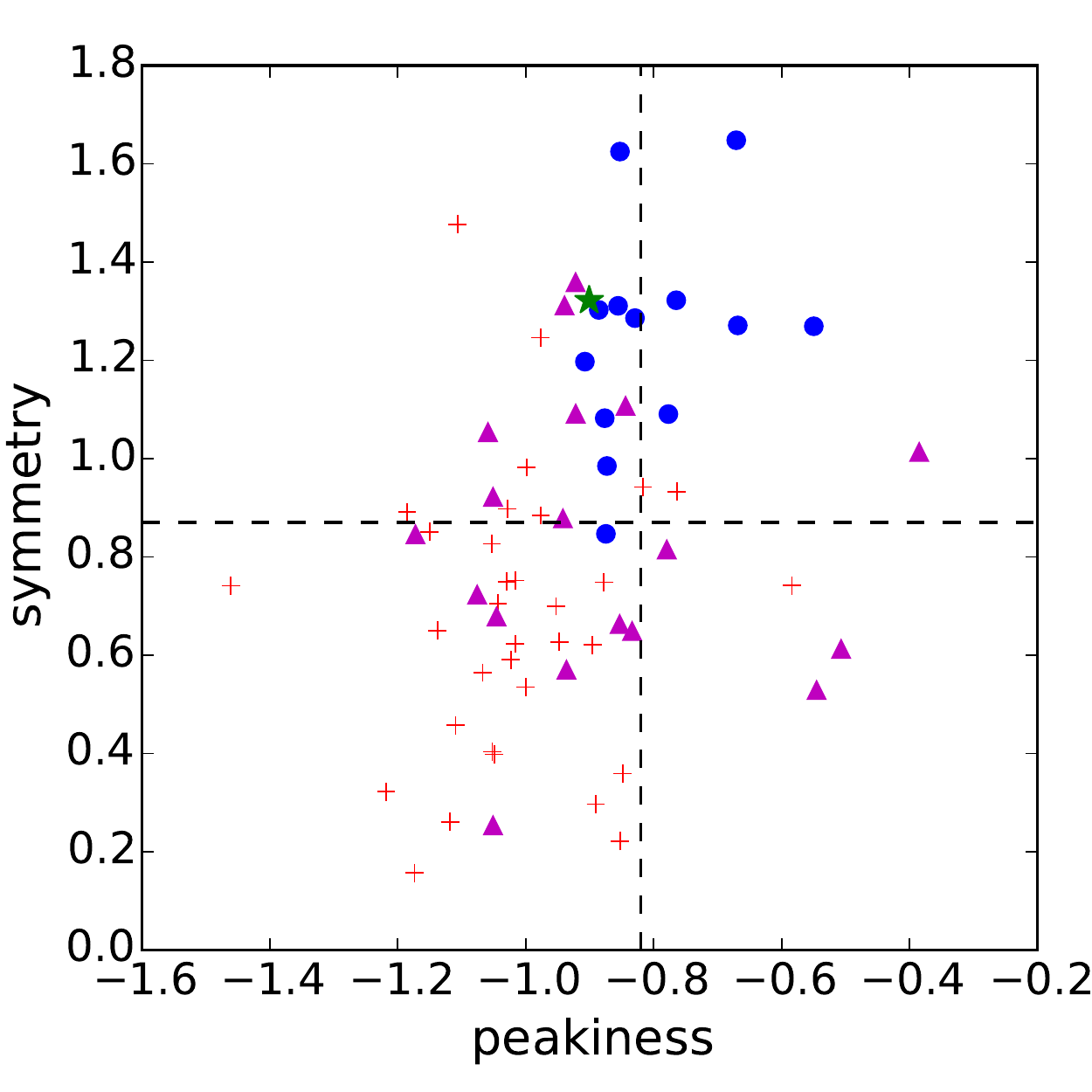}\\
    \includegraphics[width=0.33\textwidth]{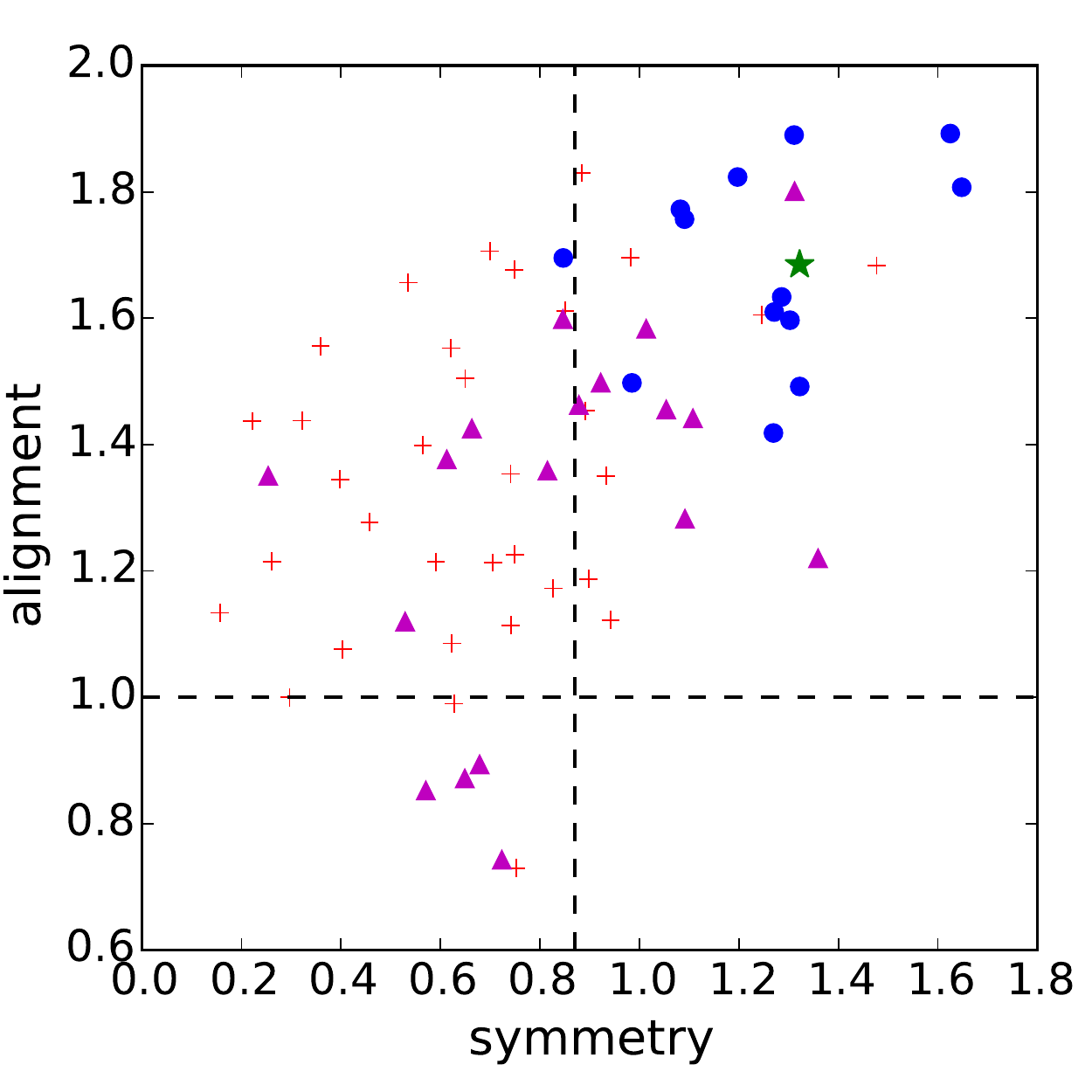}
  \end{tabular}
\caption{Distributions of \citet{mantz15b} morphological values for the
simulated mass-limited cluster sample at $z=0$. The s-p-a cuts defining the
relaxed sub-sample given in \citet{mantz15b} are shown as dashed lines. In
comparison, galaxy clusters classified into different dynamical state
categories using method given in this paper are indicated by different
markers: `very relaxed' (green star), `rather relaxed' (blue circles and the
green star), `less disturbed' (magenta triangles, blue circles and the
green star), and `disturbed' (red crosses).}
\label{fig:comparison}
\end{figure}

Recently, \citet{mantz15b} developed an automatic method for classifying the
relaxation state of galaxy clusters. Their classification is based on the
symmetry (`s'), peakiness (`p'), and alignment (`a') of their X-ray
morphologies, similar to the factors we consider for our classification
(Sect.\;\ref{sec:dynstate}). Here we compare the two classification methods on
our simulated mass-limited cluster sample at $z=0$ using mock X-ray images with 100 kilo-seconds exposure
time (Fig.\;\ref{fig:comparison}).

In \citet{mantz15b} classification, a cluster is categorized as relaxed when a
s-p-a criterion of $s > 0.87$, $p > -0.82$, and $a > 1.00$ (dashed lines in
Fig.\;\ref{fig:comparison}) is satisfied in more than 50\% of the cases
of its bootstrap analysis. This results in a relaxed cluster fraction of about
16\%, slightly smaller than the fraction (21\%) of our `rather relaxed' clusters
(blue circles and the green star in \citet{mantz15b}). In general, more relaxed
clusters in our simulated sample according to our classification also
have higher $s$-$p$-$a$ values, showing overall agreement of the two
classification methods. The specific distribution of $s$-$p$-$a$ values of our simulated
cluster sample is slightly shifted compared to the distribution of morphological values of the
\textsl{Chandra} sample (Fig.\;8, \citet{mantz15b}): our simulated clusters have
systematically higher alignment values, and lower peakiness values. This is
likely due to the inability of non-radiative simulations in reproducing
the observed properties of the cluster core. After taking the corresponding
shifted $s$-$p$-$a$ criterion into consideration, the distribution of
$s$-$p$-$a$ values of our `rather relaxed' clusters aligns rather well with
that of the relaxed clusters as classified with \citet{mantz15b} method.

\end{document}